\documentstyle[12pt,epsfig]{article}
\title{ Radiative Leptonic Decays of $D_s^{\pm} $ and $D^\pm $ Mesons }

\author{Cai-Dian  L\"{u}$^{1,2}$,Ge-Liang Song$^2$\\
{\small 1 CCAST (World Laboratory), P.O. Box 8730,
Beijing 100080, PR China}\\
{\small 2 Institute of High Energy Physics, CAS, P.O. Box 918(4), Beijing 100039, PR China}\\
\\
 }

\date{}
\begin{document}

\maketitle

\begin{picture}(0,0)(0,0)
\put(340,290){BIHEP-TH-2002-61}
\end{picture}

\begin{abstract}
In this work, we investigate the radiative leptonic decays
$D^{-}_{(s)}\to\gamma\ell \bar\nu$ ($ \ell=e, \mu$) at tree level
within the non-relativistic constituent quark model and the
effective Lagrangian for the heavy flavor decays. We find that the
contributions from  the three Feynman diagrams are all important.
With the full calculation, the decay branching ratios are of the
order of $10^{-5}$ for $D_{s}^{-}\to\gamma\ell \bar\nu$ ($ \ell=e,
\mu$) and $10^{-6}$ for $D^{-}\to\gamma\ell \bar\nu$ ($ \ell=e,
\mu$), respectively. These decays can be measured at B factories
and future CLEO-C experiments to determine the decay constants
$f_{D_s}$ and $f_D$.

\end{abstract}
\newpage

\section{Introduction}

The pure-leptonic decays of heavy mesons are useful to determine
the meson decay constants, and they are also sensitive to new
physics beyond the Standard Model(SM) \cite{dsd}. But it is well
known that the decays of $D^{-}_{(s)}$ into light lepton pairs
(see, Fig.1) are helicity suppressed by $m_\ell^2/m_{D_{(s)}}^2$:
\begin{equation}
\Gamma (D \to \ell \bar \nu )=\frac{G_F^2}{8\pi} |V_{CKM}|^2
f_{D}^2 m_{D}^3 \frac{m_\ell^2}{m_{D}^2} \left( 1-
\frac{m_\ell^2}{m_{D}^2} \right) ^2.
\end{equation}
Here $V_{CKM}$ is the corresponding Cabibbo-Kobayashi-Maskawa matrix element. In the
case of $D_s^{-}$($D^-$) decay, it is $V_{cs}$($V_{cd}$).
Fortunately the helicity suppression can be overcome by a photon
radiated from the charged particles at the cost of the
electromagnetic suppression with coupling constant $\alpha$. It is
possible that the radiative decays may be comparable or even
larger than the corresponding pure leptonic decays \cite{7}.

Several years ago, D. Atwood et al. calculated
$B^\pm(D^\pm_{s})\to\gamma\ell \nu$ in a non-relativistic quark
model \cite{ee}, with a large branching ratio. In fact, they only
considered one dominant diagram, neglecting other diagrams. They
found the order of $10^{-4}$ for the branching ratio of
$D^\pm_{s}\to\gamma\ell \nu$ decay.

Later on, Gregory P. Korchemsky et al. calculated these decays in
perturbative QCD approach \cite{8}. They gave larger branching
ratios. And C. Q. Geng et al. did the calculation in the light
front quark model \cite{tw}. Their branching ratios are rather
smaller.

In this Letter, We will study the radiative leptonic decays
$D^{-}_{(s)}\to\gamma\ell \bar\nu$ ($ \ell=e, \mu$) carefully at
tree level, using the non-relativistic constituent quark model,
similar to Ref.\cite{ee}, but including all diagrams. In the
following section, we will calculate the processes
$D^{-}_{(s)}\to\gamma\ell \bar\nu$ ($ \ell=e, \mu$) in the
framework of the constituent quark model (see, for example
\cite{hyc}). In the third section, we will compare the result with
some of the previous calculations\cite{ee,8,tw}. At last we will
conclude the calculation briefly.

\section{Model calculations }

We begin with the quark diagram calculation of $D_s^{-}\to\gamma\ell \bar\nu$ ($
\ell=e, \mu$). There are four charged particle lines in Fig.1, which correspond to
four Feynman diagrams contributing to the
radiative decays $D_s^{-}\to\gamma\ell \bar\nu$ ($ \ell=e, \mu$)
at tree level, as shown in Fig.2. However when the photon line is
attached to the internal charged line of W boson such as Fig.2d,
there is a suppression factor of $m_c^2/m_W^2$. Thus we neglect it
for simplicity. To be consistent in the following calculation, we
will always neglect the terms suppressed by the factor
$m_c^2/m_W^2$.

The decay amplitudes corresponding to the other three diagrams are
\begin{eqnarray}
{\cal H}_{a+b} &=& -i \sqrt{2} G_F e V_{cs} ~\bar c \left [  Q_c
\not \! \epsilon_\gamma \frac{ \not \!p _\gamma -\not \! p_c
+m_c}{(p_c \cdot p_\gamma)} \gamma_\mu P_L+ Q_s P_R \gamma_\mu
\frac{\not \! p_s -\not \! p_\gamma  +m_s}{(p_s\cdot p_\gamma)}
\not \! \epsilon_\gamma \right] s ~(\bar \ell \gamma ^\mu P_L \nu)
,
\nonumber\\
{\cal H}_{c} &=& -i \sqrt{2} G_F e  V_{cs} (\bar c \gamma ^\mu P_L
s) ~ \left [ \bar \ell \not \! \epsilon_\gamma \frac{ \not \!p
_\gamma +\not \! p_\ell +m_\ell }{(p_\ell \cdot p_\gamma)}
\gamma_\mu P_L \nu \right] . \label{h4}
\end{eqnarray}

As mentioned in the Introduction, we will use the constituent
quark model to reduce the amplitudes into the `hadronic level'. In
this simple model, both of the quark and anti-quark inside the
meson move with the same velocity. Thus we have
\begin{equation}
p_c^\mu=(m_c /m_{D_s})p_{D_s}^\mu, ~~~~~ p_s^\mu=(m_s
/m_{D_s})p_{D_s}^\mu. \label{s2}
\end{equation}
We use further the interpolating field technique \cite{hg} which
relate the hadronic matrix elements to the decay constants of the
mesons. The decay constant $f_{P}$ for a charged pseudoscalar meson
is defined by \cite{pdg}:
\begin{equation}
<0|A_\mu(0)|P(q)> = i f_{P} q_\mu,\label{dd}
\end{equation}
In the case of $D_s^{-}$, we have
\begin{equation}
<0|\bar c \gamma^\mu \gamma_5 s|D_s> = i f_{D_s}
p_{D_s}^\mu,\label{dd2}
\end{equation}

The whole decay amplitude for $D_{s}^{-}\to\gamma\ell \bar\nu$ ($
\ell=e, \mu$) is derived from eqs.(\ref{h4},\ref{s2},\ref{dd2}) by
neglecting the terms suppressed by $m_l/m_c$\footnote{By
neglecting terms proportional to $m_l/m_c$, we drop the infrared
divergence terms, which should be canceled by the radiative
corrections of the pure leptonic decay $D_{s}^{-}\to\ell \bar\nu$
\cite{9}.}
\begin{eqnarray}
{\cal A} &=& \frac{\sqrt{2} e G_FV_{cs}}{6(p_{D_s} \cdot
p_\gamma)}
 f_{D_s}
\left[ \left(\frac{m_{D_s}}{m_s}-2\frac{m_{D_s}}{m_c}\right) i
 \epsilon_{\mu\nu  \alpha \beta } p_{D_s}^\nu p_\gamma^\alpha
\epsilon_\gamma^\beta\right. \nonumber \\
&&\left.+\left(6- \frac{m_{D_s}}{m_s}-2\frac{m_{D_s}}{m_c}\right)
( p_{\gamma\nu} \epsilon_{\gamma\mu} -p_{\gamma
\mu}\epsilon_{\gamma\nu})p_{D_s}^\nu \right ] (\bar \ell \gamma
^\mu P_L \nu).\label{5}
\end{eqnarray}

In the $D_s^{-}$ rest frame, the differential decay width
\cite{pdg} is
\begin{equation}
{d\Gamma} =\frac{1}{(2\pi)^3} \frac{1}{32(M_{D_s})^3} |{\cal
A}|^2{d \hat s d \hat t} ,\label{6}
\end{equation}
Neglecting the mass of light leptons, we get the differential decay
width:
\begin{equation}
\frac{d\Gamma}{d \hat s d \hat t} =\frac{\alpha  G_F^2 |V_{cs}|^2
}{144\pi^2 } \frac{f_{D_s}^2}{m_{D_s}^3} \frac{\hat s}{
(m_{D_s}^2-\hat s)^2}\left[ x_s (m_{D_s}^2-\hat s-\hat t )^2 +x_c
\hat t^2 \right] ,\label{6p}
\end{equation}
with
\begin{equation}
x_s=\left( 3-\frac{m_{D_s}}{m_s}\right)^2,~~~~~
x_c=\left(3-2\frac{m_{D_s}}{m_c}\right)^2.     \label{99p}
\end{equation}
 The $\hat s$, $\hat
t$ are defined as $ \hat s =(p_\ell + p_\nu)^2$, $ \hat t =(p_\ell
+ p_\gamma)^2$. Integrating  eqn.(\ref{6p}) in phase space, we
obtain the decay width
\begin{equation}
\Gamma =\frac{ \alpha G_F^2 |V_{cs}|^2  }{2592\pi^2 } f_{D_s}^2
m_{D_s}^3 \left[ x_s+x_c \right]. \label{7p}
\end{equation}
Using $\alpha=1/137$,
 $m_c=1.5$ GeV, $m_{D_s}=1.97$ GeV, $|V_{cs}|=0.974$
\cite{pdg}, we get
\begin{equation}
\Gamma (D_s \to \gamma \ell \bar \nu )=2.3 \times {10}^{-17}
\times \left(\frac{f_{D_s} }{230MeV}\right)^2~{\rm GeV}. \label{9}
\end{equation}
For the lifetime $\tau(D_s)=0.5\times 10^{-12} s$ \cite{pdg}, and
the decay constant   used as $f_{D_s}=230MeV$ \cite{ee}, the
branching ratio is found to be $1.8\times {10}^{-5}$.
 From eqs.(\ref{99p},\ref{7p}), we can easily see that the decay
 width is sensitive to the decay constant $f_{D_s}^2$, and the
 constituent quark mass $m_c$ (or $m_s$). Any changes of the two
 input parameters, will result in a big change in the decay
 amplitude. Therefore the prediction of branching ratios remain
 the accuracy at the order of magnitude,  unless we can  precisely
 determinate the input parameters.

It is worthy of considering the differential spectrum for
experimental purposes. Deriving from eqn.(\ref{6p}), we obtain
\begin{equation}
  \label{eg}
  \frac{m_{D_s}}{\Gamma}\frac{d\Gamma}{dE_\gamma} =
\frac{1}{\Gamma}\frac{d\Gamma}{d\lambda_\gamma}=24 \lambda_\gamma
(1-2\lambda_\gamma),
\end{equation}
with $\lambda_\gamma = E_\gamma/m_{D_s}$. This result is the same
as Ref.\cite{ee}. We show the photon energy spectrum in Fig.3 as
the solid line. This is clearly distinct from the bremsstrahlung
photon spectrum.

The lepton energy distributions are
\begin{eqnarray}
  \label{en}
\frac{1}{\Gamma}\frac{d\Gamma}{d\lambda_\nu}&=&\frac{36}{x_s+x_c}
\left\{ x_c(1-2\lambda_\nu)\left[ 2\lambda_\nu +(1-2\lambda_\nu)
\ln(1-2\lambda_\nu)\right] \right.\\\nonumber &&~~~~+ \left. x_s
\left[ 2\lambda_\nu(3-5\lambda_\nu) +(1-2\lambda_\nu)
(3-2\lambda_\nu) \ln(1-2\lambda_\nu)\right]\right\},
\end{eqnarray}
\begin{eqnarray}
  \label{el2}
\frac{1}{\Gamma}\frac{d\Gamma}{d\lambda_\ell}&=&\frac{36}{x_s+x_c}
\left\{ x_s(1-2\lambda_\ell)\left[ 2\lambda_\ell
+(1-2\lambda_\ell) \ln(1-2\lambda_\ell)\right]  \right.\\\nonumber
&&~~~~+  \left. x_c \left[ 2\lambda_\ell(3-5\lambda_\ell)
+(1-2\lambda_\ell) (3-2\lambda_\ell)
\ln(1-2\lambda_\ell)\right]\right\},
\end{eqnarray}
where  $\lambda_\nu = E_\nu/m_{D_s}$, $\lambda_\ell =
E_\ell/m_{D_s}$. We show the    neutrino energy spectrum
$\frac{1}{\Gamma}\frac{d\Gamma}{d\lambda_\nu}$, and the charged
lepton (e, $\mu$) energy spectrum
$\frac{1}{\Gamma}\frac{d\Gamma}{d\lambda_\ell}$ in Fig.3 as dashed
and dash-dotted lines, respectively. Eqs.(\ref{en},\ref{el2}) are
consistent with Ref.\cite{ee}, if we consider only the diagram in
Fig.2a with the photon connecting the  strange quark line like the
case in that paper.

The formulas above can be applied to the case of $D^-$ decay
i.e. $D^-\to\gamma\ell \bar\nu$ ($ \ell=e, \mu$), directly. We get the decay
width easily:
\begin{equation}
\Gamma (D^- \to \gamma \ell \bar \nu ) =\frac{ \alpha G_F^2
|V_{cd}|^2  }{2592\pi^2 } f_{D^-}^2 m_{D^-}^3 \left[ x_d+x_c
\right]. \label{7}
\end{equation}
with $$x_d=\left( 3-\frac{m_{D^-}}{m_d}\right)^2,~~~~~
x_c=\left(3-2\frac{m_{D^-}}{m_c}\right)^2.$$ Using $m_d=0.37$ GeV,
$m_{D^-}=1.87$ GeV, $|V_{cd} |=0.22$ \cite{pdg}, we get
\begin{equation}
\Gamma (D^- \to \gamma \ell \bar \nu )=2.9 \times {10}^{-18}
\times \left(\frac{f_{D^-} }{230MeV}\right)^2~{\rm GeV}. \label{9p}
\end{equation}
For   $\tau(D^-)=1.05\times 10^{-12} s$ \cite{pdg}, the   decay
branching ratio is 4.6$\times$${10}^{-6}$ with the decay constant
$f_{D^-}=230MeV$ \cite{tw}.
Again, without the precise determination of the constituent quark
mass $m_d$, the decay branching ratio is only meaningful at the order of
magnitude.

In the case of $D^-$ decay, the formulas for the
  differential spectra of photon and lepton energy distribution
are the same as the $D_s$ decay, except replacing $x_s$ with $x_d$
in eqs.(\ref{eg},\ref{en},\ref{el2}).
 Their numerical results  are shown in
Fig.4 as solid, dashed and dash-dotted lines for $\gamma,\nu,\ell$
respectively. From Fig.3 and Fig.4, we can see that the
differential spectra of $D_s^{\pm} $ and $D^{\pm} $ radiative
leptonic decays are very similar. Only the endpoints of leptonic
energy spectra are different.


\section{Comparison with other calculations}

In Ref.\cite{ee}, D. Atwood et. al. made the calculation within
the non-relativistic constituent quark model like us. As stated in
the introduction part that they just considered the contribution
of the emission of photon from the strange quark, i.e. Fig.2a, for
they made an analogy with $B^-$ decay directly. After our careful
calculation, we conclude that the contribution of the Feynman
diagram where the photon is emitted from the initial light quark
is dominant enough to neglect the other diagrams in the case of
$B^-$ decay, but not for the case of $D_s^- $ or $ D^-$. It can be
seen at Table.1 that the contribution of the other two diagrams
corresponding to Fig.2b and 2c must be considered because   the
interference among the three Feynman diagrams is large and
destructive. That is the reason why their branching ratio of
  Br($D_{s}^-\to\gamma\ell \bar\nu$)($ \ell=e, \mu$)  decay
\cite{ee} is about four times of ours.

\begin{table}
\caption{Decay width with different diagrams and their relative
size. Considering only the diagram where the photon is emitted
from the initial light quark (Fig.2a), heavy quark (Fig.2b) or
lepton (Fig.2c), we get $\Gamma_a$, $\Gamma_b$, $\Gamma_c$,
respectively. And considering the three diagrams together, we get
$\Gamma_{a+b+c}$. }
\begin{tabular}{|c|c|c|c|c|c|}\hline
& $\Gamma_a$&$\Gamma_b$&$\Gamma_c$&$\Gamma_{a+b+c}$&$\Gamma_a$ : $\Gamma_b$ : $\Gamma_c$
: $\Gamma_{a+b+c}$ \\
\hline
 $B^-$ & 1.7$\times$${10}^{-18}$ & 5.7$\times$${10}^{-22}$ &5$\times$${10}^{-20}$ & 1.2$
 \times$${10}^{-18}$ & 1.40 : 0.0005 : 0.04 : 1 \\ \hline
 $D_s$ & 3.4$\times$${10}^{-16}$ & 8$\times$${10}^{-17}$ &4$\times$${10}^{-16}$ & 2.3$
 \times$${10}^{-17}$ & 14.72 : 3.47 : 17.32 : 1 \\ \hline
 $D^-$ & 2.1$\times$${10}^{-17}$ & 2.7$\times$${10}^{-18}$ &1.8$\times$${10}^{-17}$
 & 2.9$\times$${10}^{-18}$ & 7.30 : 0.94 : 6.03 : 1  \\ \hline
\end{tabular}
\end{table}

Gregory P. Korchemsky et al. used the perturbative QCD method to
calculate B and D meson radiative decays. Their result for $D^-$
decay is very large. In fact, the perturbative QCD approach
\cite{pqcd} is good for the B meson decays since the energy
release is very large there, but may not be good for the lighter D
meson decays. In addition, out result is consistent with that of
\cite{tw} within the light front quark model in the case of
$D_{s}^-\to\gamma e \bar\nu$. It is instructive to calculate with
various models, and the accuracy of various models will be tested
in future experiments.

\section{Summary}

We have calculated $D^{-}_{(s)}\to\gamma\ell \bar\nu$ ($ \ell=e,
\mu$) decay in non-relativistic constituent quark model. We included all the three
Feynman diagrams, and found that none of them is small in
$D_{(s)}$ decays.
We obtained the decay branching ratios of
$D_{s}^-\to\gamma\ell \bar\nu$ , $D^-\to\gamma\ell \bar\nu$($
\ell=e, \mu$) are of order ${10}^{-5}$ and ${10}^{-6}$
respectively. Such a branching ratio for the radiative leptonic
decays can be measured in the two B factories and the future CLEO-C
Experiments.

Eqs.(10,15) indicate that the decay rate of
$D^{-}_{(s)}\to\gamma\ell \bar\nu$ ($ \ell=e, \mu$) is
proportional to $f_{D_{(s)}}^2$, so one can use it to determine
the decay constant $f_{D_{(s)} }$. On the other hand, it is seen
that these processes can also be used to test the $|V_{cs}|$ and
$|V_{cd}|$ if $f_{D_s}$ and $f_{D}$ are known.

\section*{Acknowledgment}
This work is partly supported  by National
Science Foundation of China with contract No.~90103013 and 10135060.

\begin{figure}[b]
 \begin{center}
 \begin{picture}(300,200)(50,450)
 \put(0,0){\epsfxsize150mm\epsfbox{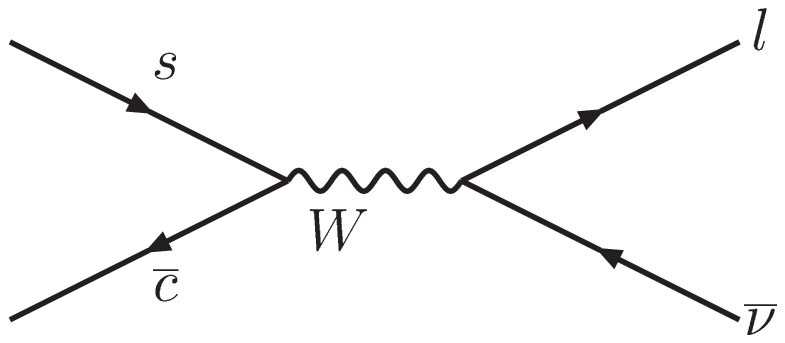}}
 \end{picture}
 \end{center}
 \caption{Feynman diagram in standard model for
$D_{s}^-$ ${\to}$ $l$ $\overline{\nu}$ decay.}
 \end{figure}

\begin{figure}[tbp]
 \begin{center}
 \begin{picture}(300,200)(160,330)
 \put(0,0){\epsfxsize200mm\epsfbox{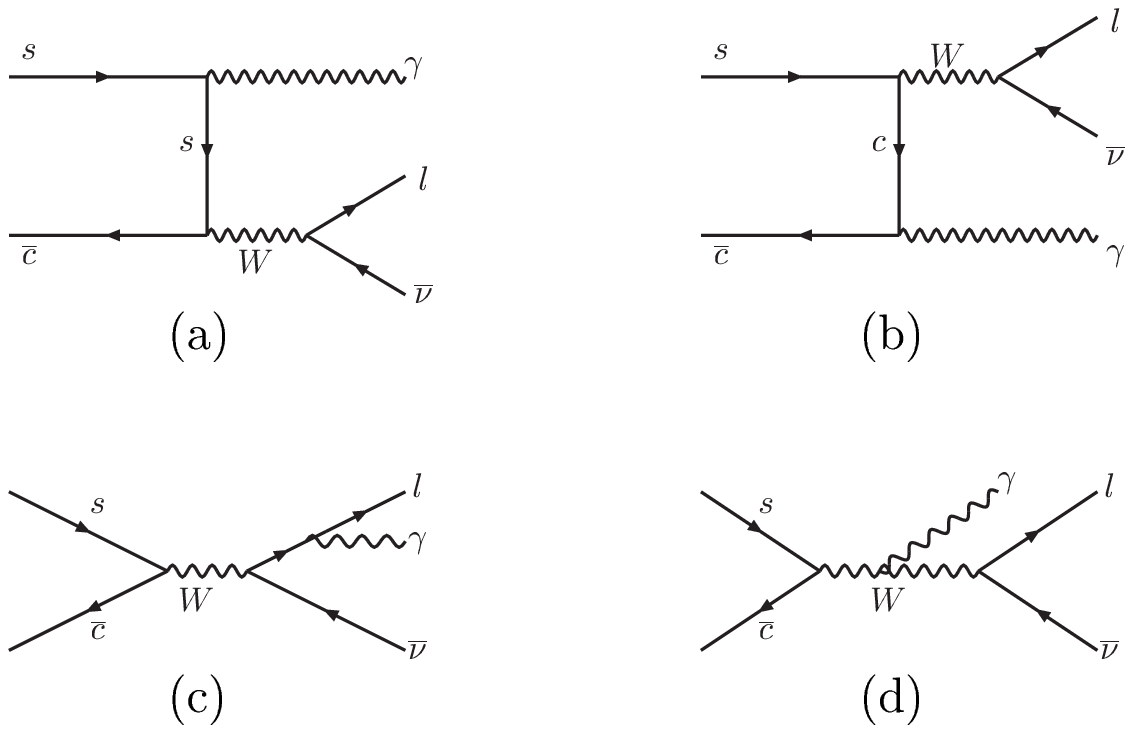}}

 \end{picture}

 \end{center}
 \caption{Feynman diagrams in standard model for
$D_{s}^-$ ${\to}$ ${\gamma}$ $l$ $\overline{\nu}$ decay.}
 \end{figure}

\begin{figure}[t]
 \begin{center}
 \begin{picture}(300,200)(0,10)
 \put(0,0){\epsfxsize110mm\epsfbox{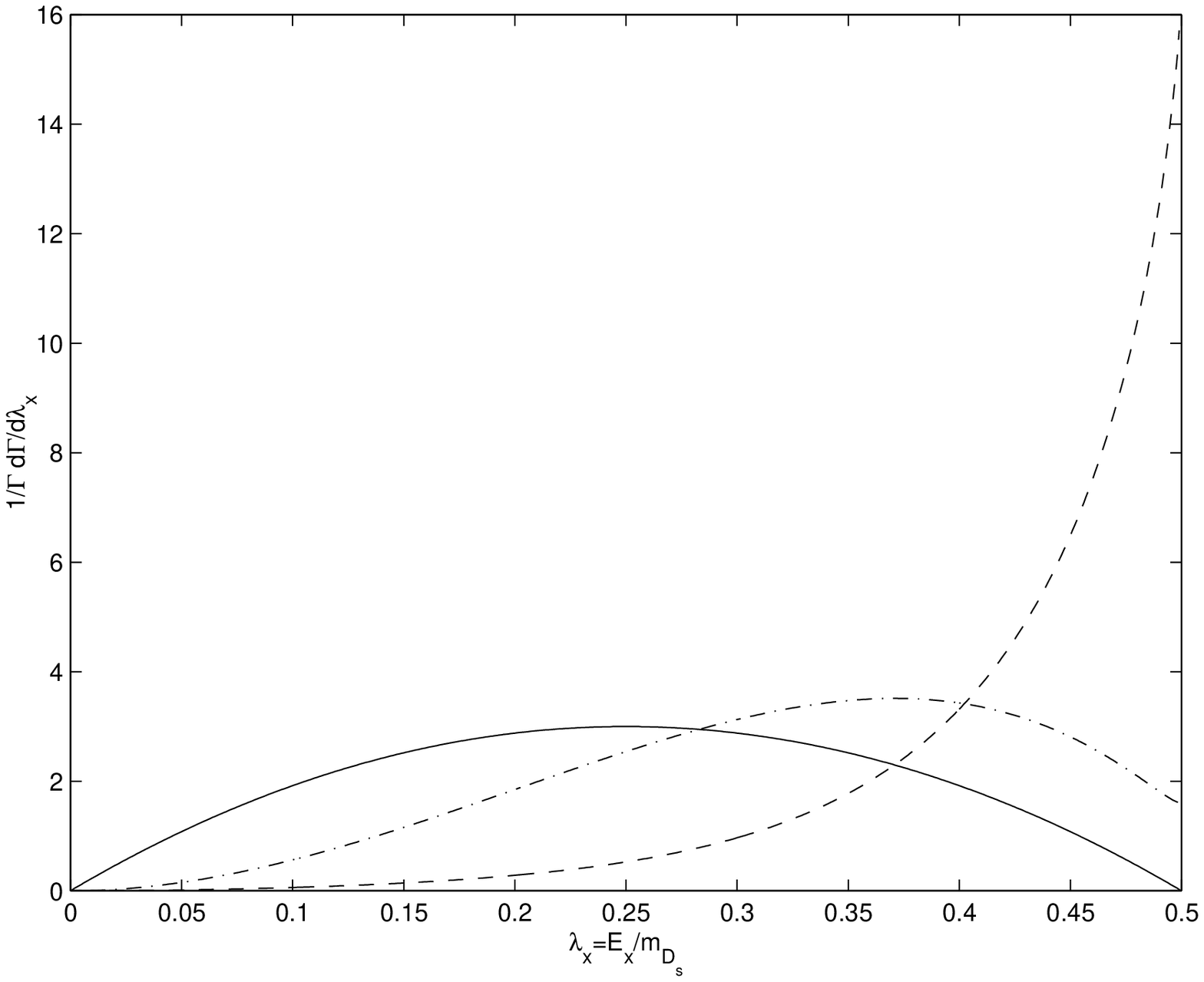}}
 \end{picture}

 \caption{Normalized energy spectra of the decay $D_s^- \to \gamma \ell \bar
\nu$. The solid line is for the photon energy spectrum, the dashed
line is for the neutrino energy and the dash-dotted line is for
the lepton energy spectrum, respectively.}
\end{center}
 \end{figure}

\begin{figure}[b]
 \begin{center}
 \begin{picture}(300,200)(0,15)
 \put(0,0){\epsfxsize110mm\epsfbox{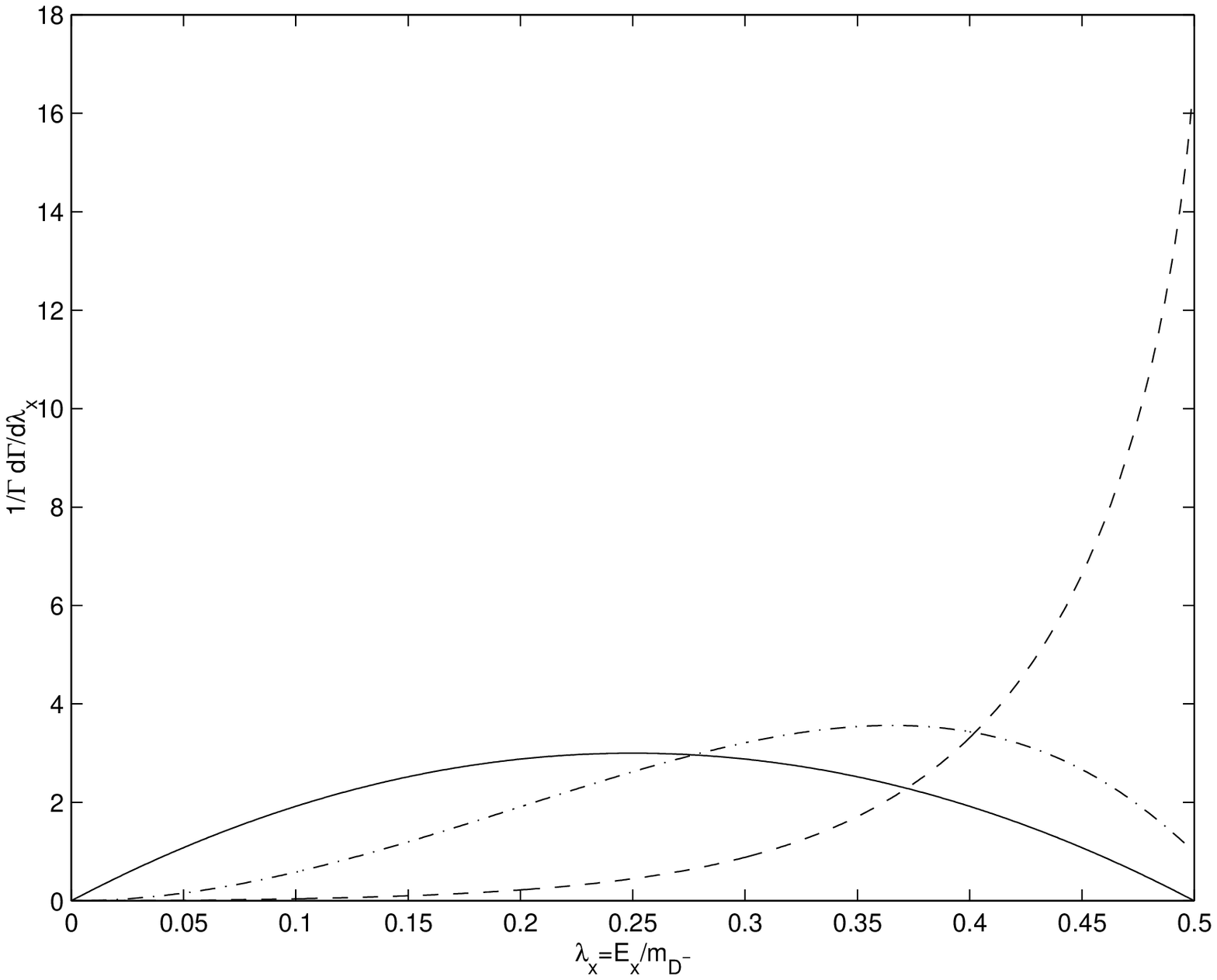}}

 \end{picture}

\caption{Normalized energy spectra of the decay $D^- \to \gamma
\ell \bar \nu$. The solid line is for the photon energy spectrum,
the dashed line is for the neutrino energy and the dash-dotted
line is for the lepton energy spectrum, respectively.}
\end{center}
 \end{figure}

\end{document}